\documentclass[10pt, paper=a4]{article}
\usepackage{graphicx}
\def\Title#1{\begin{center} {\Large #1 } \end{center}}
\def\Author#1{\begin{center}{ \sc #1} \end{center}}
\def\Address#1{\begin{center}{ \it #1} \end{center}}

\newenvironment{Abstract}{
  \begin{quotation}
  \begin{center}
  {\normalsize\bfseries ABSTRACT}
  \end{center}
  \medskip
  \noindent\ignorespaces
}{
  \end{quotation}
}

\def\Acknowledgements{\bigskip  \bigskip \begin{center} \begin{large}
      \bf ACKNOWLEDGEMENTS \end{large}\end{center}}

\newcommand{\xx}{\ensuremath{\,\text{X}}}
\usepackage{color}
\usepackage{lineno}
\usepackage{subfig}
\usepackage{hyperref}
\usepackage{amsmath}
\usepackage{tikz}
\usepackage{pgfplots}
\pgfplotsset{compat=1.16}
\usepackage{booktabs}

\def\affilone{
LPC, Clermont-Ferrand, France}

\def\affiltwo{
Department of Physics, University of Cincinnati, OH, USA}

\usepackage{fancyhdr}
\usepackage{listings}
\usepackage{xcolor}

\definecolor{functioncolor}{RGB}{0,102,204}
\definecolor{phasecolor}{RGB}{153,0,153}

\newcommand{\func}[1]{\textcolor{functioncolor}{\texttt{#1}}}

\newcommand{\phase}[1]{\textcolor{phasecolor}{\textbf{#1}}}

\begin{document}

% large size for the first page
\large
\begin{titlepage}

\vfill
\Title{Deploying a Hybrid PVFinder Algorithm for Primary Vertex Reconstruction in LHCb's GPU-Resident HLT1}
\vfill

\Author{Simon Akar$^{1}$, Mohamed Elashri$^{2,\dagger}$, Conor Henderson$^{2}$, Michael Sokoloff$^{2}$}
\Address{$^{1}$\affilone \\
$^{2}$\affiltwo \\[2pt]
$^{\dagger}$Corresponding Author}
\vfill

\begin{Abstract}
LHCb's Run 3 upgrade introduced a fully software-based trigger system operating at 30~MHz, processing an average of 5.6 proton-proton collision vertices per bunch crossing (event). This work presents the development of an inference engine for PVFinder, a hybrid deep neural network for finding primary vertices, the proton-proton collision points from which all subsequent particle decays originate into Allen, LHCb's High Level Trigger (HLT1) framework. The integration addresses critical real-time constraints including fixed memory pools, single-stream execution, and sub-400~$\mu$s per-event processing budgets on NVIDIA GPUs. We introduce a translation layer that bridges Allen's Structure-of-Arrays (SoA) data layout with  cuDNN's tensor format while maintaining zero-copy semantics and deterministic behavior. Current performance shows the CNN stage contributes significant throughput overhead. We present a roadmap targeting order-of-magnitude improvements through mixed-precision computing, model compression and other techniques.
\end{Abstract}

\vfill

\begin{quotation}
\begin{center}
{\it Contribution to the proceedings of the Connecting the Dots Workshop (CTD 2025),\\
November 10--14, 2025}
\end{center}
\end{quotation}

\vfill
\end{titlepage}
\def\thefootnote{\fnsymbol{footnote}}
\setcounter{footnote}{0}

% normal size for the rest
\normalsize 

\section{Introduction}
\label{sec:intro}

The LHCb experiment at the Large Hadron Collider \cite{LHCb_Collaboration_2008} studies matter-antimatter asymmetry through precision measurements of beauty and charm hadron decays. Run 3 operations introduced five-fold luminosity increase with respect to the previous period op operations (2010--2018), resulting in 30~MHz collision rate with 5.6 primary vertices (PVs) per event on average~\cite{lhcb_upgrade,lhcb_trigger}, with event selection performed by a fully software-based running on GPUs.

The Allen framework~\cite{allen_framework} implements LHCb's High-Level Trigger Level 1 (HLT1) on GPUs with strict constraints: sub-400~$\mu$s per-event processing, fixed memory pools, and single-stream execution. PVFinder~\cite{pvfinder_hybrid} is a novel algorithm using Machine Learning to reconstruct primary vertices in LHC collisions. The model is a hybrid deep neural network combining fully-connected (FC) layers with a one-dimensional (1D) convolutional neural networks (CNN), achieving $>$ 97\% efficiency with 0.03 false positives per event.

Achieving optimal trigger performance requires balancing speed and physics
sensitivity: improving reconstruction quality while maintaining the throughput
demanded by 30~MHz operations. Modern deep learning architectures, including
CNN, offer significant gains in physics performance over traditional heuristic approaches for tasks such as vertex finding, track fitting, and particle identification. However, integrating these algorithms into Allen's GPU-resident framework presents unique challenges, as standard ML inference patterns, dynamic memory allocation, multi-stream execution, and library-managed workspaces, conflict with Allen's deterministic, fixed-resource execution model. Resolving this tension is essential for instrumenting the trigger with state-of-the-art algorithms without compromising its real-time guarantees.

This work provides a first demonstration of how to approach this problem,
using primary vertex reconstruction as a concrete case. We present the
development of an inference engine for PVFinder inside Allen's HLT1 framework,
introducing a translation layer that bridges Allen's native CUDA environment
with cuDNN's tensor execution model for CNN inference. We characterize the
integrated throughput performance and present an optimization roadmap targeting
deployment in regular data-taking operations by 2030. Due to the nature of
this work, it should be understood as a proof-of-concept deployment that
establishes patterns applicable to future ML integration efforts in Allen.

\section{PVFinder Architecture for Real-Time Inference}
\label{sec:architecture}

PVFinder transforms reconstructed track parameters into primary vertex positions through a three-stage pipeline: feature extraction via FC layers, spatial pattern recognition using a UNet-style CNN, and heuristic peak finding. The input consists of nine features per track from VELO detector reconstruction. A fixed spatial slicing scheme divides the 400~mm interaction region into 40 intervals of 100 bins each.

The FC stage processes tracks through six FC layers implemented in native CUDA, combining each track's contribution into a 800-bin histogram (8 channels each with 100-bin). CNN stage employs a 5-layer UNet architecture (64 channels deployed): the encoder reduces spatial resolution while increasing channel depth, capturing local and contextual patterns; the decoder upsamples to original resolution producing refined probability distributions with enhanced nearby-vertex separation.

Two design decisions enable Allen compatibility: the FC stage outputs in [B, C, L] format required by cuDNN, eliminating data reorganization between stages, all memory requirements are pre-allocated from Allen's pools at initialization. The peak finding stage applies local maximum detection with learned thresholds.

\begin{figure}[!ht]
  \centering
  \includegraphics[width=0.75\columnwidth]{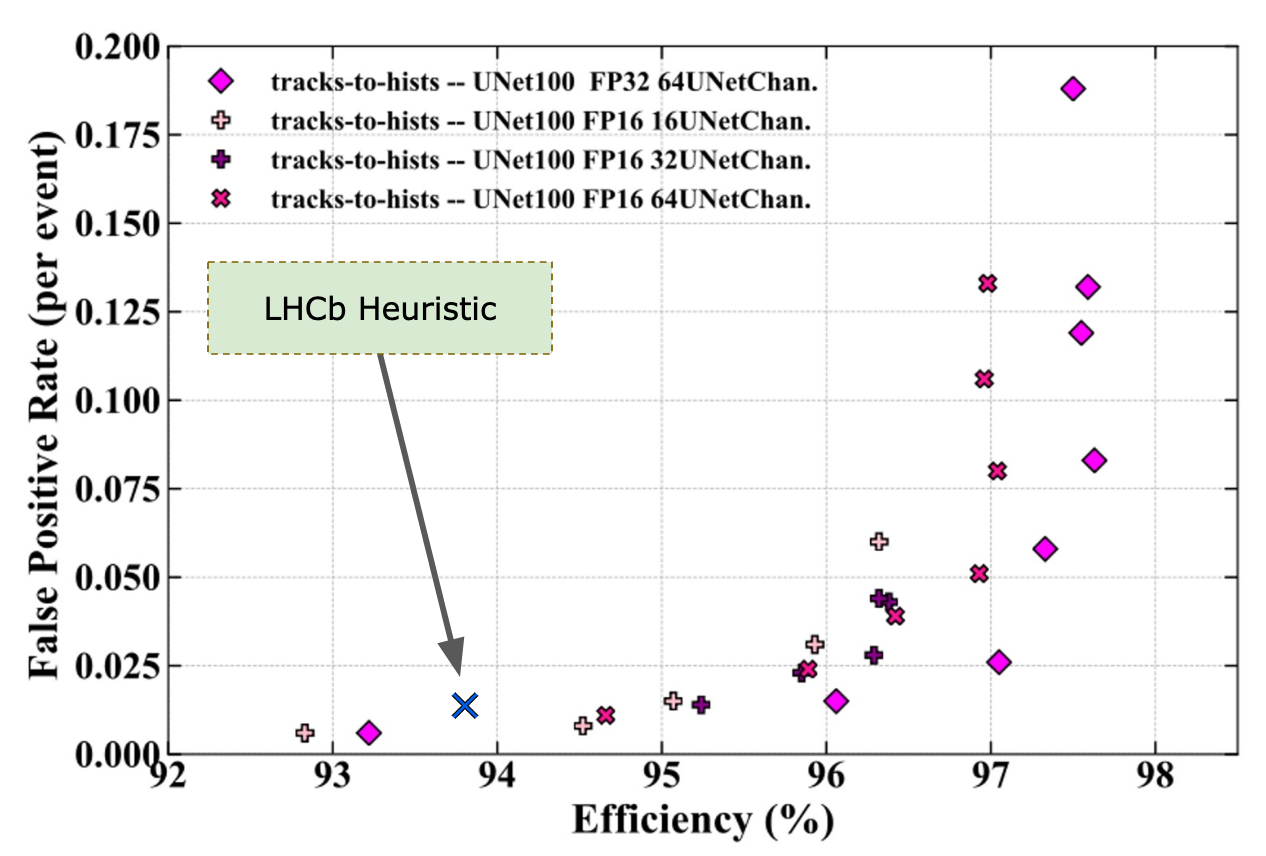}
  \caption{PVFinder physics performance showing efficiency vs. false positive rate for different configurations. The magenta configuration (FP32, 64-channel UNet) selected for deployment achieves $>$ 97\% efficiency with 0.03 false positives per event, significantly outperforming the LHCb heuristic baseline \cite{dziurda_parallel_2025}. FP16 configurations show minimal performance degradation.}
  \label{fig:physics_performance}
\end{figure}

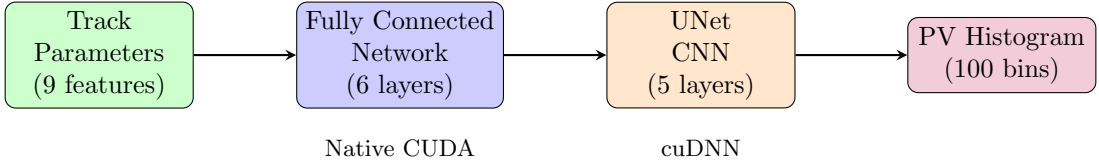
\begin{figure}[!htb]
  \centering
  \begin{tikzpicture}[
    box/.style={draw, rounded corners, minimum height=1cm, align=center},
    arrow/.style={->, thick, >=stealth}
  ]
  % Input
  \node[box, fill=green!20, minimum width=2.5cm] (input) at (0,0) {Track\\Parameters\\(9 features)};
  
  % FC Stage
  \node[box, fill=blue!20, minimum width=2.5cm] (fc) at (4,0) {Fully Connected\\Network\\(6 layers)};
  
  % CNN Stage  
  \node[box, fill=orange!20, minimum width=2.5cm] (cnn) at (8,0) {UNet\\CNN\\(5 layers)};
  
  % Output
  \node[box, fill=purple!20, minimum width=2.5cm] (output) at (12,0) {PV Histogram\\(100 bins)};
  
  % Arrows
  \draw[arrow] (input) -- (fc);
  \draw[arrow] (fc) -- (cnn);
  \draw[arrow] (cnn) -- (output);

  % Implementation details
  \node[below] at (4,-1) {\small Native CUDA};
  \node[below] at (8,-1) {\small cuDNN};
  \end{tikzpicture}
  \caption{PVFinder hybrid architecture showing the three-stage pipeline: FC layers process track parameters (9 features/track) into some representation, UNet CNN refines spatial patterns into probability histograms, and peak finding extracts vertex positions. The FC stage is implemented in native CUDA while the CNN stage uses cuDNN, bridged by the translation layer.}
  \label{fig:architecture}
\end{figure}

Physics performance on simulated events shows $>$ 97\% efficiency for 3-8 vertices per event with 0.03 false positives per event, as shown in Fig.~\ref{fig:physics_performance}. The algorithm produces 0.6\% spurious vertices per true PV. The complete architecture is shown in Fig.~\ref{fig:architecture}.

\section{Allen Integration and Translation Layer}
\label{sec:integration}

Allen's architecture imposes constraints that diverge from typical ML deployment environments. Allen processes events using stream-ordered execution where each event maps to one GPU thread block with a SoA layout for coalesced memory access. Memory management employs fixed pools, runtime dynamic allocation is forbidden to ensure predictable latency and avoid memory fragmentation that would destabilize the trigger. All operations execute on a single framework-managed stream without global synchronizations that would serialize event processing. These choices enable 30~MHz rates with deterministic performance but create interface mismatch with cuDNN's (and other standard inference libraries) NCHW tensors, workspace management, and typical multi-stream execution patterns.

The translation layer serves as a three-phase adapter implementing zero-copy semantics while maintaining deterministic guarantees. The \phase{Prepare} phase creates non-owning tensor views reinterpreting Allen's SoA buffers as cuDNN-compatible tensors. Since FC output is already in [B, C, L] format (B=40 spatial intervals, C=64 channels, L=800 bins per interval), most data requires no reorganization. We allow tracks to contribute to one or two intervals which means that the tracks on boundaries of intervals are part of two intervals and are handled as pre-allocated staging buffers. Shape validation and descriptor creation occur at initialization time; per-event overhead reduces to pointer arithmetic and bounds checking, contributing negligible latency to the critical path.

The \phase{Execute} phase invokes cuDNN convolution operations on Allen's stream, preserving event-parallel semantics without introducing serialization points. A fixed workspace arena (16~MB) sized for largest intermediate activations is pre-allocated, bound to the cuDNN handle at initialization, and reused across all events. This eliminates per-event allocation overhead that would violate Allen's memory model constraints. The implementation ensures no default CUDA stream operations leak into the execution path, which would introduce implicit synchronizations violating Allen's deterministic execution model.

The \phase{Extract} phase maps cuDNN outputs to Allen's SoA buffers for downstream algorithm consumption. Output data remains in cuDNN-native layout until explicitly accessed by consumers, minimizing unnecessary data movement and cache pollution. The peak finding algorithm reads directly from this layout, avoiding additional format conversion.

The API exposes minimal initialization (TensorSpec freezing batch dimensions, channel count, sequence length, and data type; \func{init()} method accepting Allen's stream, workspace allocation, and cuDNN backend handle) and lightweight per-event execution (TensorView creation, \func{execute()} invocation). Error handling employs a Status enum enabling graceful degradation: events failing validation bypass ML reconstruction without disrupting data acquisition.

\section{Integration Performance Results}
\label{sec:performance}

Performance evaluation focuses on characterizing throughput impact within Allen's HLT1 budget. All measurements are performed on NVIDIA RTX 2080 Ti GPUs.

Standalone measurements characterize the computational cost of each pipeline stage. The FC stage processes events at 300~kHz throughput (~5\% HLT1 budget), while the complete hybrid model reaches 110~kHz, indicating CNN dominance.

Integration into Allen's full HLT1 sequence reveals the overhead of running within the complete reconstruction chain. Baseline HLT1 serves as the reference 100\%. Adding only FC reduces to ~95\%, consistent with standalone measurements. Full hybrid implementation reduces to ~25\%—a 75\% reduction, as summarized in Table~\ref{tab:throughput}.

\begin{table}[!htb]
  \begin{center}
    \begin{tabular}{l|c|c}
      \hline
      \hline
      Configuration & Relative Throughput & Notes \\
      \hline
      Baseline HLT1 & 100\% & Reference \\
      + FC stage only & $\sim$95\% & Native CUDA \\
      + Full hybrid (FC+CNN) & $\sim$25\% & Current \\
      \hline
      Target (2030) & $\geq$95\% & After optimization \\
      \hline
      \hline
    \end{tabular}
    \caption{HLT1 throughput impact measured on NVIDIA RTX 2080 Ti. The CNN stage dominates overhead, reducing throughput to 25\% of baseline. Optimization roadmap targets return to $\geq$95\% performance.}
    \label{tab:throughput}
  \end{center}
\end{table}

Profiling identifies three contributing factors. First, memory bandwidth saturates under the combined load of VELO tracking algorithms, PVFinder inference, and downstream trigger reconstructions. Standalone measurements isolated PVFinder's bandwidth consumption and could not capture these interference effects. Second, cache interference between PVFinder's working set (model parameters, intermediate activations) and other algorithms' data structures degrades memory access latency across the entire trigger chain, creating system-wide performance degradation beyond PVFinder's direct costs. Third, the unoptimized CNN implementation does not fully utilize available compute resources; streaming multiprocessor occupancy remains below 50\% during convolution operations due to suboptimal kernel launch configurations that leave compute units idle while memory operations dominate execution time.

While current implementation does not meet $<$ 5\% regression targets, the detailed characterization validates the integration approach and identifies specific, actionable optimization paths.

\section{Optimization Roadmap and Outlook}
\label{sec:outlook}

Performance analysis reveals CNN stage improvements are required to reach $<$ 5\% HLT1 throughput regression. Multiple optimization strategies address the identified bottlenecks with clear technical justification.

Mixed-precision inference using FP16 provides ~2$\times$ throughput improvement through two mechanisms. First, halving precision from 32 to 16 bits doubles arithmetic intensity, allowing the same memory bandwidth to feed twice as many operations. Second, FP16 enables tensor core utilization on modern GPUs, providing dedicated hardware for matrix multiplication with 1.5-2$\times$ additional speedup through specialized 4$\times$4$\times$4 multiply-accumulate (MMA) units. Physics validation confirms FP16 introduces $<$0.5\% efficiency degradation, well within acceptable tolerances for trigger applications. The combination of bandwidth and compute improvements yields 3-4$\times$~\cite{Akar:2023zhd} total speedup from precision alone. 
Model compression via 32-channel UNet (versus current 64 channels) provides 4$\times$ speedup due to quadratic scaling of convolution cost with channel count. Convolution operations scale as $O(C_{in} \times C_{out} \times K)$ where $C$ denotes channels and $K$ is kernel size. Halving channels therefore reduces compute by 4$\times$ while memory footprint (model parameters and activations) also decreases by 4$\times$. This
addresses both the compute and memory bandwidth bottlenecks identified in profiling. Physics studies indicate the 32-channel variant maintains $>$96\% efficiency, suggesting the 64-channel model is over parameterized for this task. Knowledge distillation from the 64-channel teacher to 32-channel student model can potentially recover any marginal performance loss.

Memory layout optimizations target the observed cache interference and bandwidth saturation. Fusing the FC stage output directly into contiguous [B, C, L] cuDNN-native tensors eliminates intermediate format conversion overhead and improves spatial locality. Explicit workspace reuse minimizes working set size, reducing cache pollution for concurrent algorithms. Launch parameter tuning addresses the $<$50\% on average SM occupancy by optimizing thread block dimensions and register allocation, allowing better hiding of memory latency through increased parallelism. Profiling shows that current convolution kernels achieve only 40-45\% theoretical occupancy due to register pressure and shared memory conflicts. Restructuring thread block geometry to reduce per-thread register usage while maintaining sufficient parallelism can increase occupancy to 75-80\%, directly translating to improved throughput. Combined memory optimizations contribute an additional 1.5$\times$ speedup.

The optimization strategy prioritizes changes with highest impact-to-effort ratio. FP16 conversion requires minimal code changes (data type specifications and range validation) but delivers immediate 2$\times$ gains. The 32-channel model necessitates retraining but the training infrastructure already exists. Memory optimizations involve systematic kernel restructuring guided by profiler feedback. This phased approach allows incremental validation at each step rather than attempting all optimizations simultaneously.

\begin{table}[!htb]
  \centering
  \begin{tabular}{lcl}
    \toprule
    Optimization & Expected Speedup & Implementation Effort \\
    \midrule
    FP16 + Tensor Cores & $4\xx$ & Require significant code changes \\
    32-channel UNet & $4\xx$ & Infrastructure exists, testing needed \\
    Memory layout fusion & $1.5\xx$ & Kernel restructuring guided by profiler \\
    SM occupancy tuning & (included above) & Systematic launch parameter optimization \\
    \midrule
    \textbf{Combined (optimistic)} & $\boldsymbol{\sim}$\textbf{24$\xx$} & Phased approach with incremental validation \\
    \bottomrule
  \end{tabular}
  \caption{Optimization strategies with expected throughput improvements and implementation effort. FP16 conversion offers the highest impact-to-effort ratio. The 32-channel model compression addresses algorithmic complexity through retraining. Memory layout fusion and kernel tuning target cache interference and suboptimal SM occupancy ($<$50\%). Combined optimizations project sufficient speedup to reduce CNN overhead from 75\% to $\sim$3\%, meeting the $<$5\% target for 2030 deployment. These are optimistic upper-bound estimates assuming independent optimization gains.}
  \label{tab:optimization_priority}
\end{table}

Optimistic estimates project: FP16+tensor cores (4$\xx$) $\xx$ 32-channel UNet (4$\xx$) $\times$ memory optimizations (1.5$\xx$) = 24$\xx$ total speedup. This would reduce CNN overhead from 75\% reduction to ~3-5\%, well within the $<$~5\% target for 2030 deployment. Each optimization addresses specific bottlenecks identified through profiling: bandwidth (FP16), compute (tensor cores), algorithmic complexity (32-channels), and cache/occupancy (memory tuning). It is worth mentioning that these are optimistic estimations scenarios where we assume that each of these optimizations are independent which is probably not in practice so these should be taken as optimistic upper bound. The multiplicative assumption requires that optimizations don't interfere with each other. For example, FP16 may reduce register pressure, affecting the memory optimization gains.

The translation layer establishes a template for additional ML algorithms in Allen. Several groups pursue GPU-accelerated ML for track fitting, particle ID, and heavy flavor tagging. Backend generalization (\texttt{ONNX Runtime}, \texttt{TensorRT}) is under consideration after PVFinder optimization. 

\section{Conclusions}

This work presents the deployment of a hybrid deep neural network for finding PVs into LHCb's GPU-resident trigger system, demonstrating feasibility of real-time ML inference under strict deterministic constraints. The translation layer successfully bridges Allen's SoA layout with cuDNN's tensor operations while maintaining zero-copy semantics and single-stream execution. Physics validation confirms $>$ 97\% reconstruction efficiency with 0.03 false positives per event, matching standalone implementations.

Current measurements reveal the CNN stage contributes 75\% throughput overhead. The optimization roadmap combining mixed-precision computing, model compression, and memory improvements projects order-of-magnitude speedup sufficient for 2030 requirements. The integration approach establishes a foundation for deploying additional ML algorithms in LHCb's trigger system.

The experience demonstrates that successful ML deployment in real-time trigger systems requires careful co-design of data formats, memory management, and execution models. Early integration into the target framework validates architectural assumptions and identifies system-level issues that isolated development misses.

\Acknowledgements
The authors acknowledge the LHCb Real Time Analysis Project for providing the software and computing infrastructure and the Allen development team for technical guidance during integration. This work was supported by the Institute for Research and Innovation in Software for High Energy Physics (IRIS-HEP) through NSF Cooperative Agreement PHY-2323298. Disclaimer: ``Any opinions, findings, and conclusions or recommendations expressed in this material are those of the author(s) and do not necessarily reflect the views of the National Science Foundation.''

\newpage
\bibliographystyle{unsrt}
\bibliography{ref}

\end{document}